\documentclass[preprint]{aastex}
\usepackage{graphicx, amsmath}

\begin{document}

\title{Measurement of $H(z)$ and $D_A(z)$ from the two-dimensional power spectrum of Sloan Digital Sky Survey luminous red galaxies}
\author{Maddumage Don P. Hemantha, Yun Wang}
\affil{Homer L. Dodge Department of Physics \& Astronomy, University of Oklahoma, 440 W Brooks Street, Norman, OK 73019, USA}
\author{Chia-Hsun Chuang}
\affil{Instituto de F{\'i}sica Te{\'o}rica, (UAM/CSIC), Universidad Aut{\'o}noma de Madrid, Cantoblanco, E-28049 Madrid, Spain}

\begin{abstract}We present a method to measure the Hubble parameter $H(z)$ and the angular diameter distance $D_A(z)$ simultaneously from the two-dimensional matter power spectrum from galaxy surveys with broad sky coverage. We validate this method by applying it to the LasDamas mock galaxy catalogs. Then we apply this method to Sloan Digital Sky Survey (SDSS) Data Release 7 and obtain measurements of $\Omega_mh^2=0.1268 \pm 0.0085$, $H(z=0.35)=81.3\pm 3.8$km/s/Mpc, $D_A(z=0.35) = 1037\pm44$Mpc, without assuming a dark energy model or a flat universe. We also find that the derived parameters $H(0.35)r_s(z_d)/c=0.0431 \pm 0.0018$ and $D_A(0.35)/r_s(z_d)=6.48 \pm 0.25$. These are in excellent agreement with similar measurements from the two-dimensional correlation function of the same data.
\end{abstract}

\section{INTRODUCTION}
Galaxy redshift survey data are essential for contemporary precision cosmology as they provide a method to study large-scale structure of the Universe with increasing accuracy as the number of galaxies included grow exponentially. Early surveys such as Canada-France Redshift Survey(CFRS) contained only 591 galaxies (\cite{1995ApJ...455..108L}), Harvard-Smithsonian Center for Astrophysics 2 (CfA2) survey contained 19,369 galaxies (\cite{1999PASP..111..438F}), Las Campanas Redshift Survey (LCRS) consists of 26,418 redshifts of galaxies (\cite{1996ApJ...470..172S}), and Point Source Catalog redshift (PSCz) survey measured redshifts of 15,411 galaxies (\cite{2000MNRAS.317...55S}) using Infra-Red Astronomical Satellite(IRAS). Most of these are all sky surveys. Recent efforts such as the 2dF Galaxy Redshift Survey (2dFGRS) measured redshifts of 221,414 galaxies (\cite{2003astro.ph..6581C}), WiggleZ survey measured 238,770 galaxy redshifts (\cite{2012PhRvD..86j3518P}), and SDSS obtained redshift of 930,000 galaxies in the seventh data release, DR7, (\cite{2009ApJS..182..543A}). The SDSS-III Baryon Oscillation Sky Survey (BOSS) is targeting 1.5 million Luminous Red Galaxies (LRGs) (\cite{2013AJ....145...10D}) while the Euclid mission will obtain redshifts of
approximately 50 million galaxies (\cite{2009ExA....23...39C, 2010MNRAS.409..737W}).

The galaxy power spectrum is obtained through Fourier transforming the observed galaxy distribution. One dimensional power spectrum formed by spherically averaging the Fourier space has been studied well (eg: \cite{2005MNRAS.362..505C, 2001MNRAS.327.1297P, 2010MNRAS.401.2148P, 2010MNRAS.404...60R}) to estimate cosmological parameters including matter density and Hubble's constant. In our previous paper, we presented the analysis of one dimensional two point correlation function (1D2PCF) from the same data (\cite{2012MNRAS.423.1474C}). Similar studies have used different data sets such as \cite{2005ApJ...633..560E}, \cite{2009MNRAS.393.1183C}, and \cite{2010ApJ...710.1444K}. However, it is not possible to measure both $H(z)$ and $D_A(z)$ from one dimensional power spectrum or 2PCF alone. The first simultaneous measurement of both of these quantities was obtained by \cite{2012MNRAS.426..226C} using the SDSS DR7 two-dimensional two point correlation function (2D2PCF). Although the power spectrum and the 2PCF are a Fourier pair, they provide information complementary to each other as redshift surveys cover a limited volume of the Universe. Therefore, we analyze two-dimensional galaxy power spectrum in this study.

The two-dimensional galaxy power spectrum has been studied from different redshift surveys: Las Campanas survey (\cite{1996ApJ...456L...1L}), WiggleZ survey (\cite{2010MNRAS.406..803B} and \cite{2011MNRAS.415.2876B}), HETDEX project (\cite{2012AAS...21942414C}), for example. However, the estimation of the full set of cosmological parameters was not carried out. \cite{2001MNRAS.325.1389J} measured 2D galaxy power spectrum for $0.25\le k \le 2.5h$Mpc$^{-1}$ using LCRS data. However, their limited data set prevented them from measuring the complete set of cosmological parameters. \cite{2003PhRvD..68f3004H} explored the possibility of extracting the Hubble parameter, $H(z)$, and angular diameter distance, $D_A(z)$, from future surveys and noted that curvature of the sky needs to be handled correctly for a broad sky survey such as SDSS. The WiggleZ data was used to obtain 2D power spectrum and estimate bias and growth rate as well as cosmic expansion rate at several redshifts (\cite{2010MNRAS.406..803B,2011MNRAS.415.2876B}, \cite{2011MNRAS.418.1725B}). However, the underlying cosmological model used throughout that analysis was fixed to \emph{Wilkinson Microwave Anisotropy Probe} (WMAP) best fit parameters. Our study aims to measure the main cosmological parameters in addition to $H(z)$ and $D_A(z)$ from the two-dimensional power spectrum.

In section \ref{data}, we describe the data set used. The method used to obtain the two-dimensional power spectrum is presented in section \ref{methodology}. In section \ref{results}, we validate our method using simulated data and then present the results obtained from real data. We also compare the parameter values with similar work in section  \ref{results} and summarize our findings in section \ref{discussion}.

\section{DATA}\label{data}
The SDSS-II project was finished in October 2008 and this final public data release included spectroscopic observations of 9380 square degrees of sky. These observations were carried out with 2.5 m telescope (\cite{2006AJ....131.2332G}) at Apache Point Observatory in New Mexico, United States. The luminous red galaxy (LRG) sample (\cite{2001AJ....122.2267E}) used in this work was extracted from \texttt{dr72full0} the New York University-Value Added Galaxy Catalog (NYU-VAGC) (\cite{2005AJ....129.2562B}) by setting the flag \texttt{primTarget = 32}. The K-correction was applied to NYU-VAGC data assuming a $\Lambda$CDM fiducial model with $\Omega_m = 0.3, h = 1$. We have selected LRGs located within the redshift range $0.16-0.47$ and excluded Southern Galactic Cap region, resulting in an LRG sample of 89,599.

Spectra of individual galaxies are obtained by placing fibres on the focal plane of the telescope to guide the light from individual objects to spectrometers. The finite size of these fibres makes it impossible to measure galaxies closer than 55'', a problem known as ``fibre collisions". Although the overlapping of spectroscopic tiles (\cite{2003AJ....125.2276B}) alleviates this issue partially through multiple observations, some galaxies in crowded regions were not observed. \cite{2002ApJ...571..172Z} showed that assigning the redshift of the nearest galaxy with measured redshift is sufficient for large scale structure studies. VAGC used this procedure to correct for fibre collisions.

The angular selection function is generated from the geometry and completeness information provided by VAGC in terms of spherical polygons. We have used the MANGLE (\cite{2008MNRAS.387.1391S}) software package to apply the angular selection function to the data and random galaxies. The radial selection function was constructed by binning the galaxy sample with redshift bins of size $\Delta z = 0.01$.

\section{METHODOLOGY}\label{methodology}
\subsection{2D Galaxy Power Spectrum Estimation}
\begin{figure}[!h]
    \centering
    \includegraphics[width=6in]{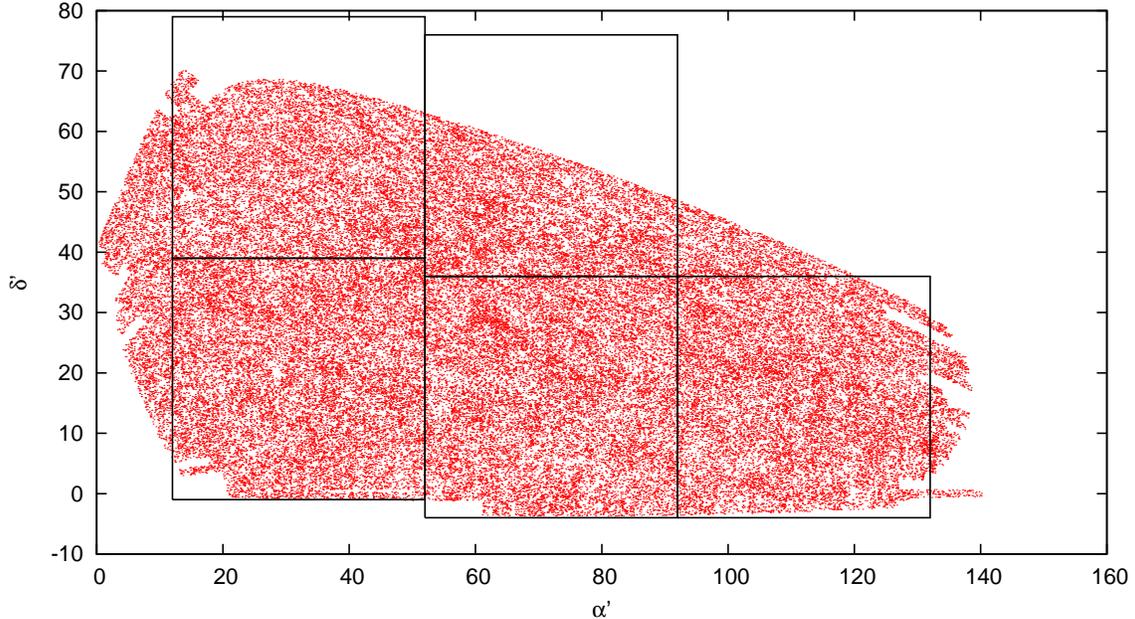}
    \caption{This is a plot of SDSS DR7 LRG galaxy sample using a Sanson-Flamsteed projection. The five patches we use are shown. Note that the coordinates are not equatorial (RA, Dec). From left to right, patches 1-3 are the lower panels, and patches 4 and 5 are the upper panels. }\label{tiles40x40}
\end{figure}

In this section, we describe the power spectrum estimation method, which is a two-dimensional extension of the FKP estimator (\cite{1994ApJ...426...23F}). The first step is tiling  the SDSS sky coverage into equal area patches as shown in Fig.\ref{tiles40x40}. This is necessary as the flat sky approximation will not hold for a survey with extended sky coverage such as SDSS. We used the Sanson-Flamsteed projection (\cite{2012psa..book.....W}) where a given Right Ascension ($\alpha$), Declination ($\delta$) pair is mapped such that,
\begin{equation}
\alpha^\prime = \alpha \cos{\delta}, \qquad \delta^\prime = \delta
\end{equation}
to generate equal area patches.

Choosing too small patches decreases the number of galaxies inside each patch, thus increasing the shotnoise.
Choosing patches that are too big will lead to deviation from the flat sky approximation. We have tested dividing the entire survey area into 2, 5, and 10 patches. We find that the 5 patch division yields the lowest bias on estimated parameters, based on application to the SDSS DR7 LRG mocks from the LasDamas (Large suit of Dark matter simulations) collaboration
(McBride et al., in preparation) (see section \ref{results_validation} for further details). Therefore, we divide the sky into five patches throughout this paper. Galaxies inside each patch were converted to a cartesian coordinate system such that x axis is pointed towards the center of each patch. Distances to galaxies are calculated from redshifts assuming a $\Lambda$CDM fiducial model (the same as used by LasDamas in making the LRG mocks) with matter density fraction, $\Omega_m = 0.25$. Each patch is then Fourier transformed as described below. Our choice of axes means that $k_{\|} = k_x$ and $k_\bot = \sqrt{k_y^2 + k_z^2}$.

We enclose each patch individually in a cube of side 2000$h^{-1}$ Mpc, and use the Nearest Grid Point (NGP) scheme (\cite{1988csup.book.....H}) to interpolate weighted galaxy positions to a regular grid of size 512$^3$. We use the standard FKP optimal weights (for minimum variance), $w(\mathbf{r})=\bar{n}(\mathbf{r})/(1+\bar{n}(\mathbf{r})\bar{P}),$ where $\bar{n}(\mathbf{r})$ is the expected number density of galaxies and $\bar{P} = 10000 h^{-3}$Mpc$^3$ is the average amplitude of the power spectrum. We tested the robustness of this choice by using $\bar{P} = 40000h^{-3}$Mpc$^3$ instead, and verified that the exact value of $\bar{P}$ has virtually no effect on the shape of the power spectrum. The FKP estimator described in Eq.2.1.3 of FKP is calculated at each grid point, and then the fast Fourier transform of the grid was obtained. A random galaxy set was generated using MANGLE with the same sky coverage and angular selection function as the real LRG sample. We have used approximately one hundred times more random galaxies than real LRGs to minimize the shot noise. The random galaxies are also divided into the same five patches described above before being used. The Fourier space was then cylindrically summed with bin size $\Delta k = 0.01h$Mpc$^{-1}$ in each direction and the shot noise term is subtracted to obtain 2D power spectrum with $z$ axis pointed in $k_{\|}$ direction. We retain only the region $0.02h$ Mpc$^{-1} \le k \le 0.16h$ Mpc$^{-1}$ where $k = \sqrt{k_{\|}^2 + k_{\bot}^2}$ to minimize the effects from aliasing (\cite{2005ApJ...620..559J}).

\subsection{Theoretical Model}
A theoretical model power spectrum is necessary for extracting cosmological parameters from the measured 2D power spectrum. We use the model,
\begin{equation}P^\textrm{s}_\textrm{dw}(k, \mu, z_0) = P_\textrm{dw}(k, \mu, z_0)\frac{(1 + \beta\mu^2)^2}{1 + (k\mu\sigma_v)^2}\label{model1}\end{equation} (\cite{1987MNRAS.227....1K,1994MNRAS.267.1020P,1998ASSL..231.....H}), where $\beta$ is the redshift distortion parameter, $\sigma_v$ is the pairwise peculiar velocity dispersion divided by $H_0$, and $\mu$ is the cosine of the angle between the line of sight and wave vector $\mathbf{k}$. $P_\textrm{dw}(k, \mu, z_0)$ is the dewiggled linear galaxy power spectrum given by,
\begin{equation}
P_\textrm{dw}(k, \mu, z_0) = G^2(z_0) P_\textrm{0} k^{n_s} T_\textrm{dw}^2(k, \mu, z_0),
\end{equation}
where $G(z_0)$ is the linear growth factor and $n_s$ is the power-law index of the primordial matter power spectrum. Anisotropicaly dewiggled transfer function, $T_\textrm{dw}(k, \mu, z_0)$, is constructed from the linear transfer function, $T_\textrm{lin}(k, z_0)$, and the ``no wiggle'' transfer function, $T_\textrm{nw}(k, z_0)$ from Eq.(29) of \cite{1998ApJ...496..605E} as in \cite{2013MNRAS.430.2446W},
\begin{equation}
T^2_\textrm{dw}(k, \mu, z_0) = T^2_\textrm{lin}(k, z_0)\exp{(-g_\mu k^2 / k^2_{\star})} + T^2_\textrm{nw}(k, z_0)(1 - \exp{(-g_\mu k^2 / k^2_{\star})}),
\end{equation}
where $g_\mu$ is given by
\begin{equation}
g_\mu = G^2(z_0)[1 - \mu^2 + \mu^2(1 + f_g^2(z_0))^2]
\end{equation}

We use $z_0 = 0.35$ as the average redshift in this paper, following previous work on the same data. We use CAMB (\cite{2000ApJ...538..473L}) to calculate linear transfer functions. For the efficient calculation of $T_\textrm{lin}(k, z_0)$ for parameters ($\Omega_bh^2, \Omega_ch^2$), where $\Omega_b$ and $\Omega_c$ are the baryon and dark matter density fractions
respectively, and $h$ is the dimensionless Hubble constant ($H_0=100h$km/s/Mpc), we create an evenly spaced grid of transfer functions with spacing 0.001 and 0.005 respectively in each parameter. Cubic spline interpolation is then used to find the linear theory transfer function for a given set of parameter values. This process is much faster than running CAMB and was rigorously tested and found to be accurate for fitting purposes in this paper. However, linear theory power spectrum does not adequately describe the galaxy power spectrum due to non linear effects. We use  a modified version (\cite{2008MNRAS.390.1470S}) of the semi-analytic model introduced by \cite{2005MNRAS.362..505C} to correct the linear matter power spectrum, and modify the galaxy power spectrum as follows:
\begin{equation}
P^\textrm{s}_\textrm{nl} = \frac{1 + Qk^2}{1 + Ak + Bk^2}P^\textrm{s}_\textrm{dw}(k, \mu, z_0),
\label{pk2dmodel}
\end{equation}
where, $A, B, Q$ are constants. Following \cite{2008MNRAS.390.1470S}, we fix $B = Q / 10$ and this seem to fit the observed galaxy power spectrum on the range of interest ($0.02h$Mpc$^{-1}\le |\mathbf{k}| \le 0.16h$Mpc$^{-1}$).

\begin{figure}[!h]
    \begin{tabular}{ll}
        \includegraphics[width=2.5in, angle = -90]{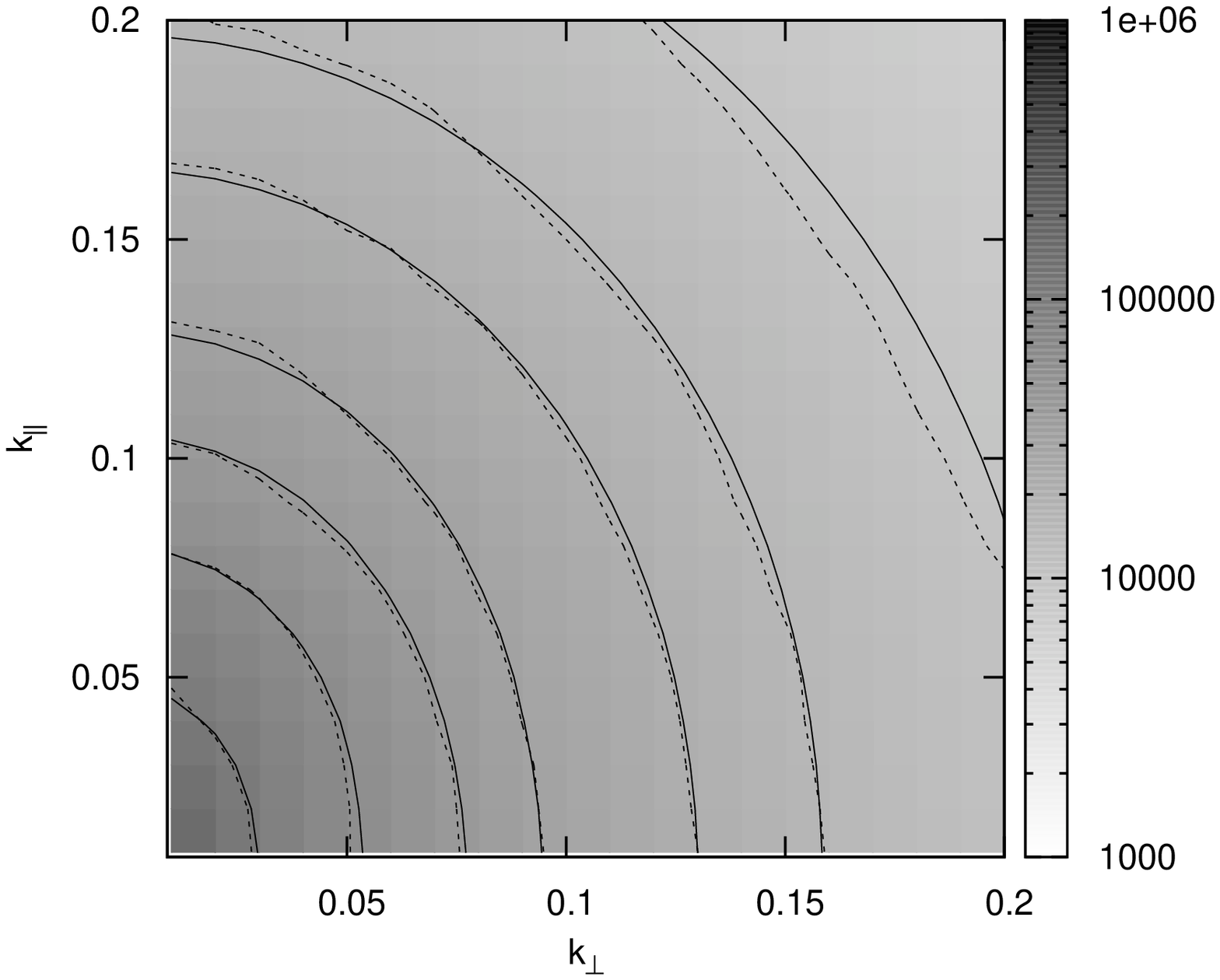}&
        \includegraphics[width=2.5in, angle = -90]{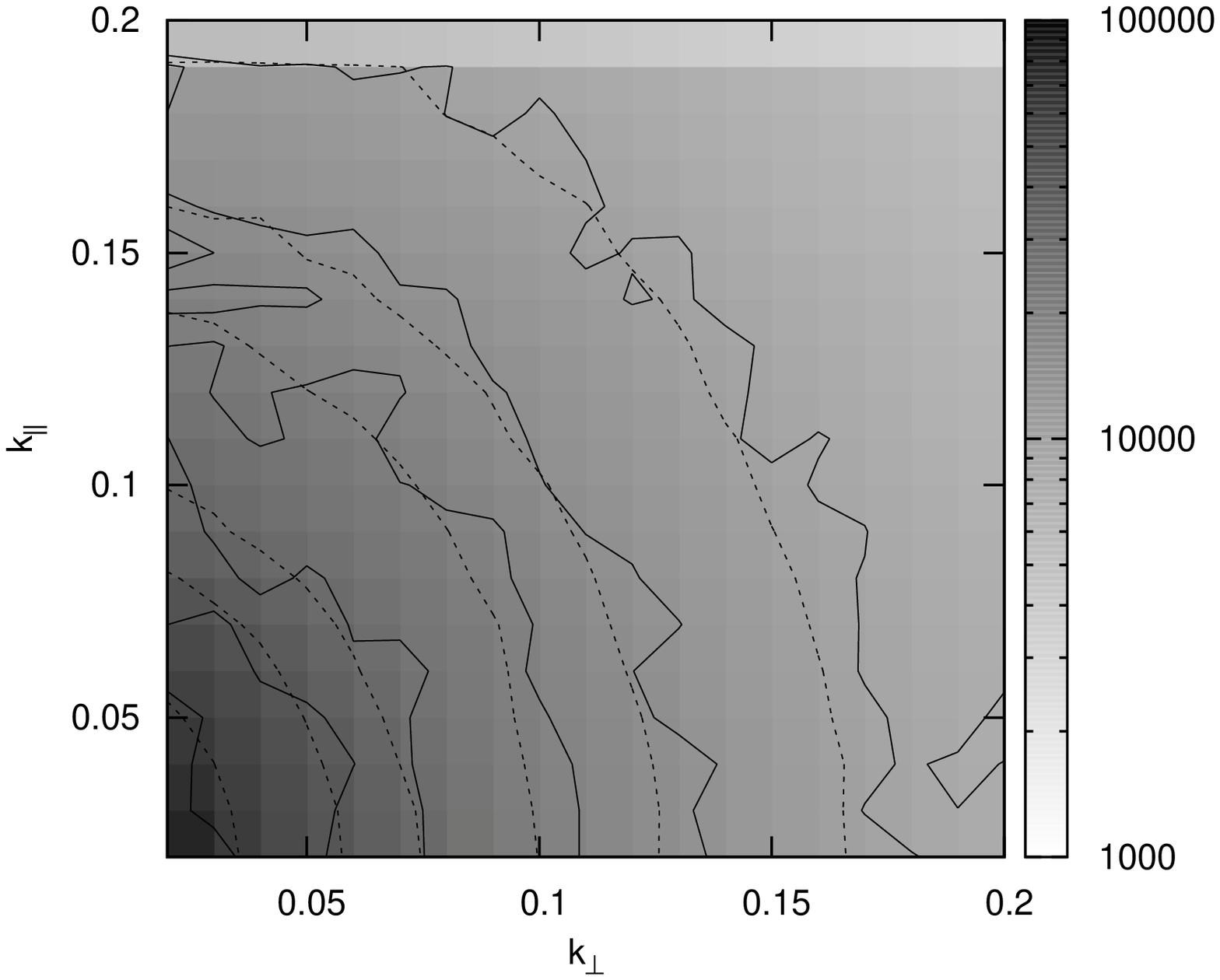}\\
    \end{tabular}
    \caption{Left: Comparison of the average of 160 LasDamas 2D galaxy power spectra (solid lines) and our model 2D power spectrum convolved with the appropriate window (dotted lines). Model parameters are set to the LasDamas input values. Contour levels are in log scale. Right: Average 2D power spectrum from SDSS DR7 LRGs (solid lines). All five power spectra from different patches were averaged to obtain a smooth plot. The best fit model corresponding to the parameters listed in Table \ref{sdssresults}, convolved with window functions of five patches and averaged together, is plotted with dashed lines.}\label{ldmodelcompare}
\end{figure}

Fig.\ref{ldmodelcompare} (left panel) shows a comparison of our theoretical model and the average of 2D power spectra obtained from 160 LasDamas mock catalogs. As discussed in the next section, the model spectrum is convolved with the window function of each of the five patches and then averaged to obtain a smooth plot. This shows the non-linear correction model is able to approximate the observed galaxy power spectrum within our range of interest.

\subsection{Window Matrix}
The observed galaxy power spectrum, $P_\textrm{obs}(\mathbf{k})$, is given by convolving the
true galaxy power spectrum, $P_\textrm{t}(\mathbf{k})$, with the survey window function, $W(\mathbf{k})$, as follows:
\begin{equation}
P_\textrm{obs}(\mathbf{k}) = \int d^3\mathbf{k\prime}P_\textrm{t}(\mathbf{k^\prime})|W(\mathbf{k - k^\prime})|^2,\label{convolution1}
\end{equation}
where the window function is given by
\begin{equation}
W(\mathbf{k}) = \int d^3\mathbf{r}\bar{n}(\mathbf{r})w(\mathbf{r})\exp(i\mathbf{k\cdot r}).
\end{equation}

As cylindrical coordinate system is a natural choice for 2D power spectrum, Eq.(\ref{convolution1}) can be rewritten as,
\begin{equation}
P_\textrm{obs}(\mathbf{k}) = \int dk_\|^\prime dk_\bot^\prime d\phi^\prime k_\bot^\prime P_t(\mathbf{k^\prime}) |W(\mathbf{k - k^\prime})|^2.
\label{convolution}
\end{equation}

The survey window function in configuration space, $w(\mathbf{r})$, is obtained from the random galaxy catalog by using NGP scheme on weighted random catalog alone on the previously mentioned $512^3$ size grid. In theory, one can deconvolve the observed power spectrum with the window function to obtain the underlying true galaxy power spectrum. However, deconvolution is susceptible to noise degradation. Thus, we convolve the model with the window window function instead, and compared the convolved model
with the observed galaxy power spectrum.

Starting with a cube of size 8000$\,$Mpc$h^{-1}$ and successively dividing the size by a factor of 2 until the size is 500$\,$Mpc$h^{-1}$ (similar to \cite{2005MNRAS.362..505C}), we construct a full three dimensional survey window by only keeping the range 25\% - 50\% of Nyquist frequency from each box. We use periodic boundary conditions to map points that lie outside boxes. It is necessary to use multiple boxes to obtain a window function with sufficiently wide range ($0.0004\,h$Mpc$^{-1} \le |\mathbf{k}| \le 0.7979\,h$Mpc$^{-1}$).
We repeat this procedure for each of our five patches, and obtain five window functions. As the convolution process given by Eq.(\ref{convolution}) is numerically expensive, we do this integration one time and cast the result into a window matrix $W_{i,j}$. $P_t(\mathbf{k})$ is replaced by a set of unit basis vectors and the contribution of the window is calculated on each basis vector. For a fixed set of $i\equiv(k_\|,k_\bot)$ and $ j\equiv(k^\prime_\|,k^\prime_\bot)$,
\begin{equation}
W_{i,j} = k_\bot^\prime\int_0^{2\pi}d\phi|W(\mathbf{k - k^\prime})|^2.
\end{equation}

The window matrix terms are normalized such that $\sum_j W(i,j) = 1$ for each $i$. Pre calculated 3D window is spline-interpolated(\cite{1992nrfa.book.....P}) to carry out the integration. Now, using Eq.(\ref{convolution}), a 2D model galaxy power spectrum given by Eq.(\ref{pk2dmodel}) can be convolved with the SDSS window function as follows:
\begin{equation}
P_{\textrm{th},i} = \sum_j P_{\textrm{gal},j}W_{i,j}.
\end{equation}

We construct window matrices for each patch separately, and convolve each with the model 2D power spectrum,
to obtain the model power spectrum for each patch. The model for each patch can be compared with the observed power spectrum of that patch
in a likelihood analysis.

\subsection{Covariance Matrix}

\begin{figure}[!h]
    \begin{tabular}{ll}
        \includegraphics[width=2.5in, angle = -90]{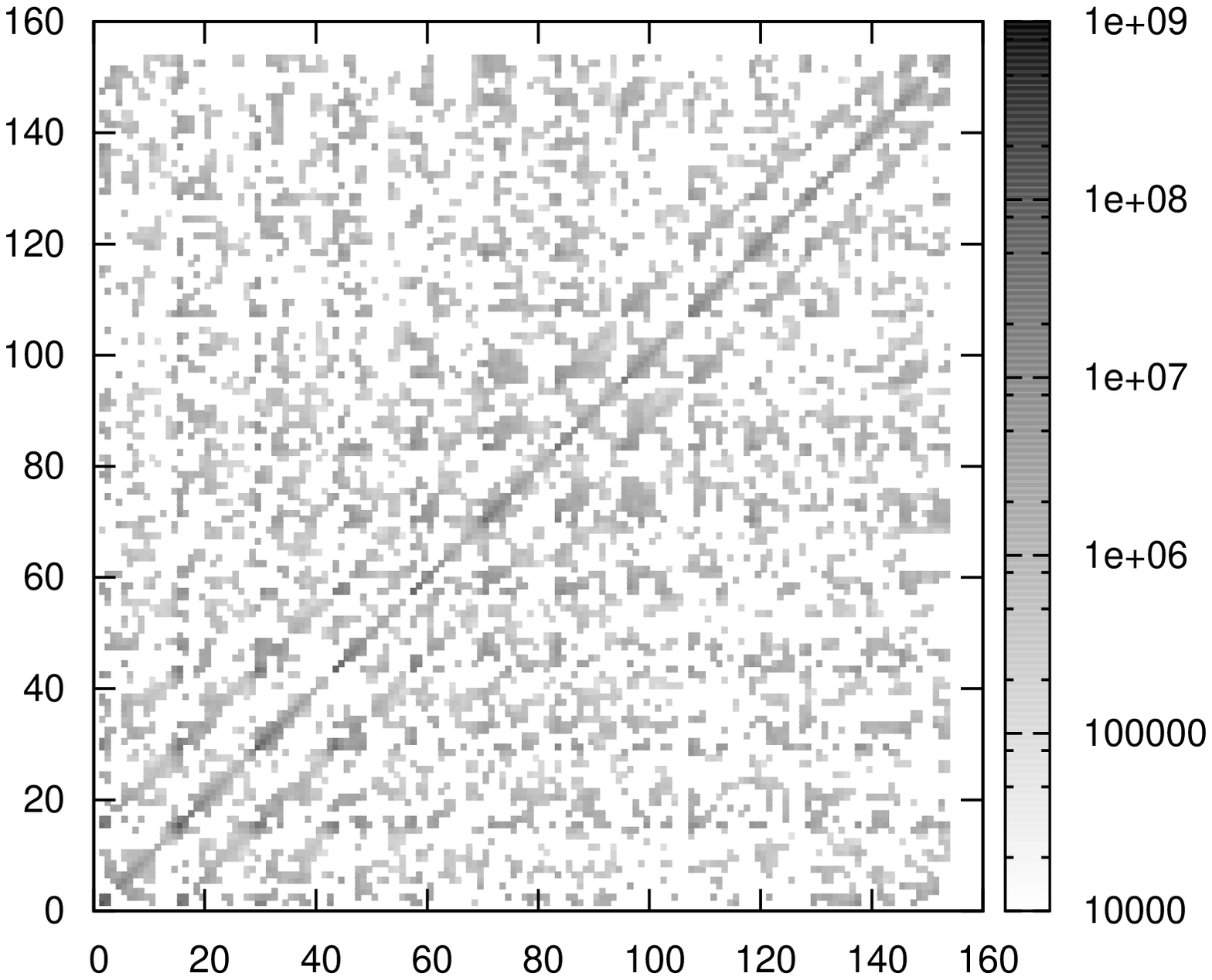} &
        \includegraphics[width=2.5in, angle = -90]{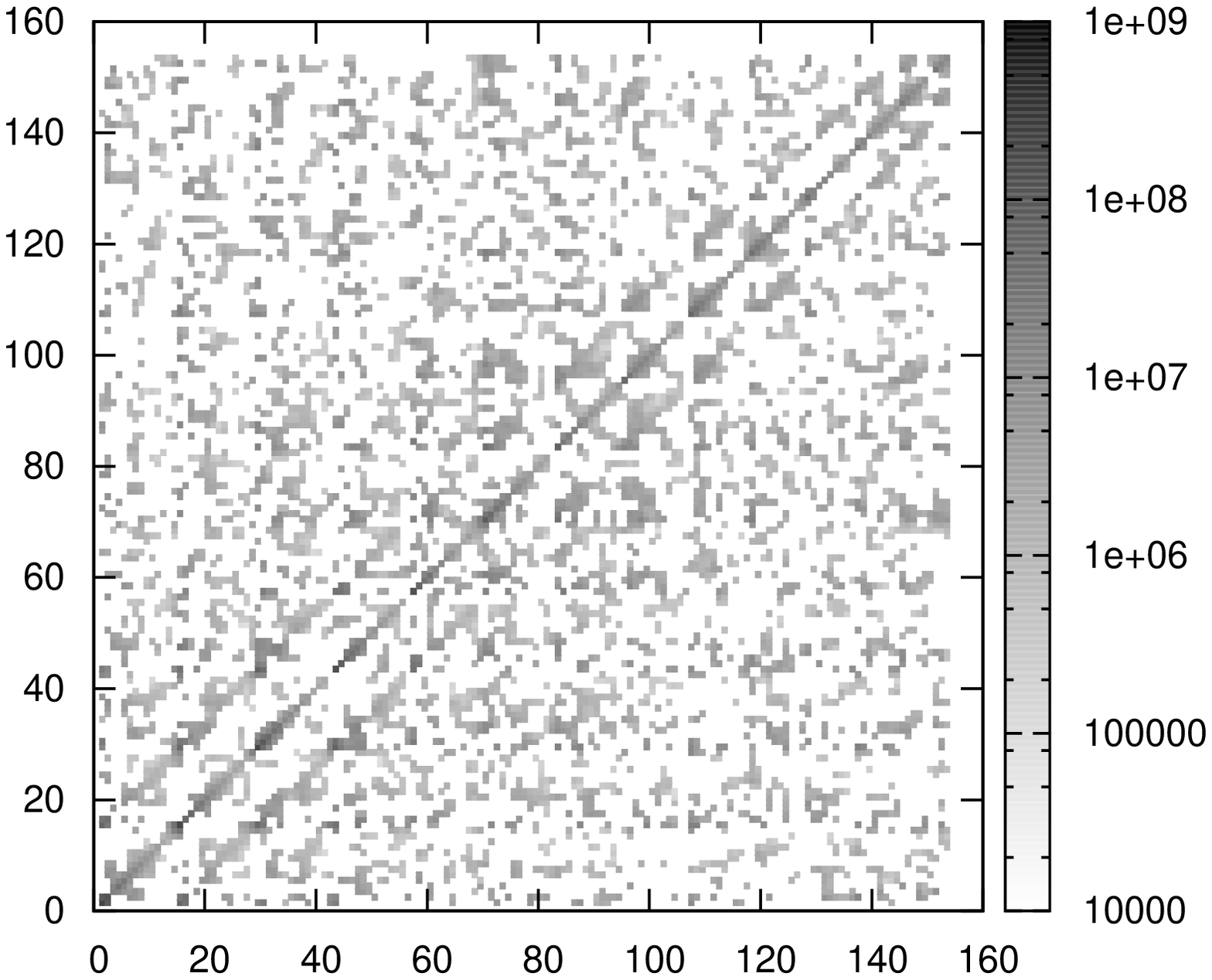}\\
    \end{tabular}
    \caption{Covariance matrices for SDSS data set(left) and LasDamas mock data(right). Both covariance matrices are calculated for the same patch. These matrices are created by unrolling the actual 2D array of points inside the area of interest where there are 154 points.}\label{covmat}
\end{figure}

We estimate the covariance matrix as follows
\begin{equation}
C_{ij} = \frac{1}{N-1}\sum_k (\bar{P_i} - P_i^k)(\bar{P_j} - P_j^k),
\end{equation}
where $N$ is the number of mocks catalogs, $\bar{P_i}$ is the mean power spectrum at the $i$th bin, and $P_i^k$ is the power spectrum at the $i$th bin in the $k$th mock catalog. We construct a total of five covariance matrices (one each for the five patches shown in the Fig.\ref{covmat}). For convenience, we unroll the 2D array of points inside the mask $0.02\le |\mathbf{k}| \le 0.16$ and construct a 1D array of 154 points. This allows us to express the covariance matrix as a 2D matrix.

We use 160 LasDamas mocks to generate covariance matrix for SDSS data. As the galaxy density of the volume limited LasDamas mocks are different from luminosity limited SDSS real galaxy sample, we dilute the mock catalog using the rejection method so that both SDSS and mock data have the same radial selection function. These covariance matrices need smoothing due to the fact that there are only 160 mock catalogs available, and the diluting process described above further reduces the number of galaxies by about 20\% in each catalog. We use the same method as described in \cite{2012MNRAS.426..226C} to make covariance matrices smooth. We use their Eq.(A1) with $p = 0.01$, $\Delta s=\Delta k=0.01h$Mpc$^{-1}$ and repeat the process ten times. The diagonal elements are smoothed using their
Eq.(A2) with the same parameter choices.

\subsection{Likelihood}

We derive constraints on estimated parameters in a Markov Chain Monte Carlo (MCMC) likelihood analysis.
The likelihood is proportional to $\exp{(-\chi^2)}$, with
\begin{equation}
\chi^2 = \sum_{i,j} (P_{\textrm{obs},i} - P_{\textrm{th},i})C_{ij}^{-1}(P_{\textrm{obs},j} - P_{\textrm{th},j}).
\label{chisqrd}
\end{equation}
To use this equation in its original form, one needs to recalculate the covariance matrix and the observed 2D power spectrum
for each set of cosmological parameters under consideration (e.g.,\cite{2005MNRAS.362..505C}).
We use the scaling method from \cite{2012MNRAS.426..226C}, which has the advantage that
the observed 2D power spectrum and its covariance matrix only need to be calculated once.
The scaling operator $\mathrm{T}$ is defined as,
\begin{equation}
P_\textrm{obs}(k_\|,k_\bot) = \mathrm{T}(P_\textrm{obs}^\textrm{fid}(k_\|,k_\bot)),
\end{equation}
where $P_\textrm{obs}^\textrm{fid}(k_\|,k_\bot)$ is the observed power spectrum obtained using a fiducial cosmological model for distance estimation. Now, Eq.(\ref{chisqrd}) can be written as,
\begin{equation}
\chi^2 = \sum_{i,j}[\mathrm{T}^{-1}(P_{\textrm{th}, i}) - P_{\textrm{obs}, i}^\textrm{fid}]
C_{\textrm{fid},ij}^{-1}[\mathrm{T}^{-1}(P_{\textrm{th}, j}) - P_{\textrm{obs}, j}^\textrm{fid}].
\label{eq:chi2}
\end{equation}

The scaling operator $\mathrm{T}$ can be constructed by considering the size of an object of observed size $\Delta z, \Delta \theta$ in the line of sight and transverse directions respectively. Using this argument, \cite{2003ApJ...598..720S} found,
\begin{equation}
k_\bot^\textrm{fid} = k_\bot \frac{D_A(z)}{D_A^\textrm{fid}(z)};\; k_\|^\textrm{fid} = k_\| \frac{H^\textrm{fid}(z)}{H(z)}.
\end{equation}
We define our scaling operator using the above relations, and apply it to the theoretical power spectrum as follows,
\begin{equation}
\mathrm{T}^{-1}(P_\textrm{th}(k_\|,k_\bot)) = P_\textrm{th}\left(\frac{D_A^\textrm{fid}(z)}{D_A(z)}k_\bot, \frac{H(z)}{H^\textrm{fid}(z)}k_\|\right),
\end{equation}
which we use to calculate $\exp{(-\chi^2)}$ (see Eq.[\ref{eq:chi2}]).

We use COSMOMC (\cite{2002PhRvD..66j3511L}), a publicly available package for MCMC likelihood analysis. Cosmological parameters $\Omega_bh^2$ and $n_s$ are fixed at WMAP 7 values as these parameters are not well constrained by power spectrum alone, and $k_\star = 0.11h$Mpc$^{-1}$ is used as results are found to be insensitive to small changes of $k_\star$. We use the data to extract constraints on $\{\Omega_mh^2$, $H(0.35)/H^\textrm{fid}(0.35)$, $D_A^\textrm{fid}(0.35)/D_A(0.35)\}$, and marginalize over parameters $\{\beta$, $\sigma_v$, $Q$, $A$, $N\}$ where $N$ is the normalization of the power spectrum. We use flat priors $\beta = [0.0, 0.9]$, $\sigma_v = [0.0, 700.0]$km/s, $Q = [5.0, 30.0]h^{1/2}$Mpc$^{-1/2}$, and $A = [0.5, 10.0]h$Mpc throughout this work.

\section{RESULTS}\label{results}

We will first present the results from appying our method to mocks (which establish the validity
of our method), then the results from the analysis of SDSS DR7 LRGs.

\subsection{Validating the Method Using Mock Data}\label{results_validation}

\begin{figure}
    \begin{tabular}{ll}
        \includegraphics[width=2in, angle = -90]{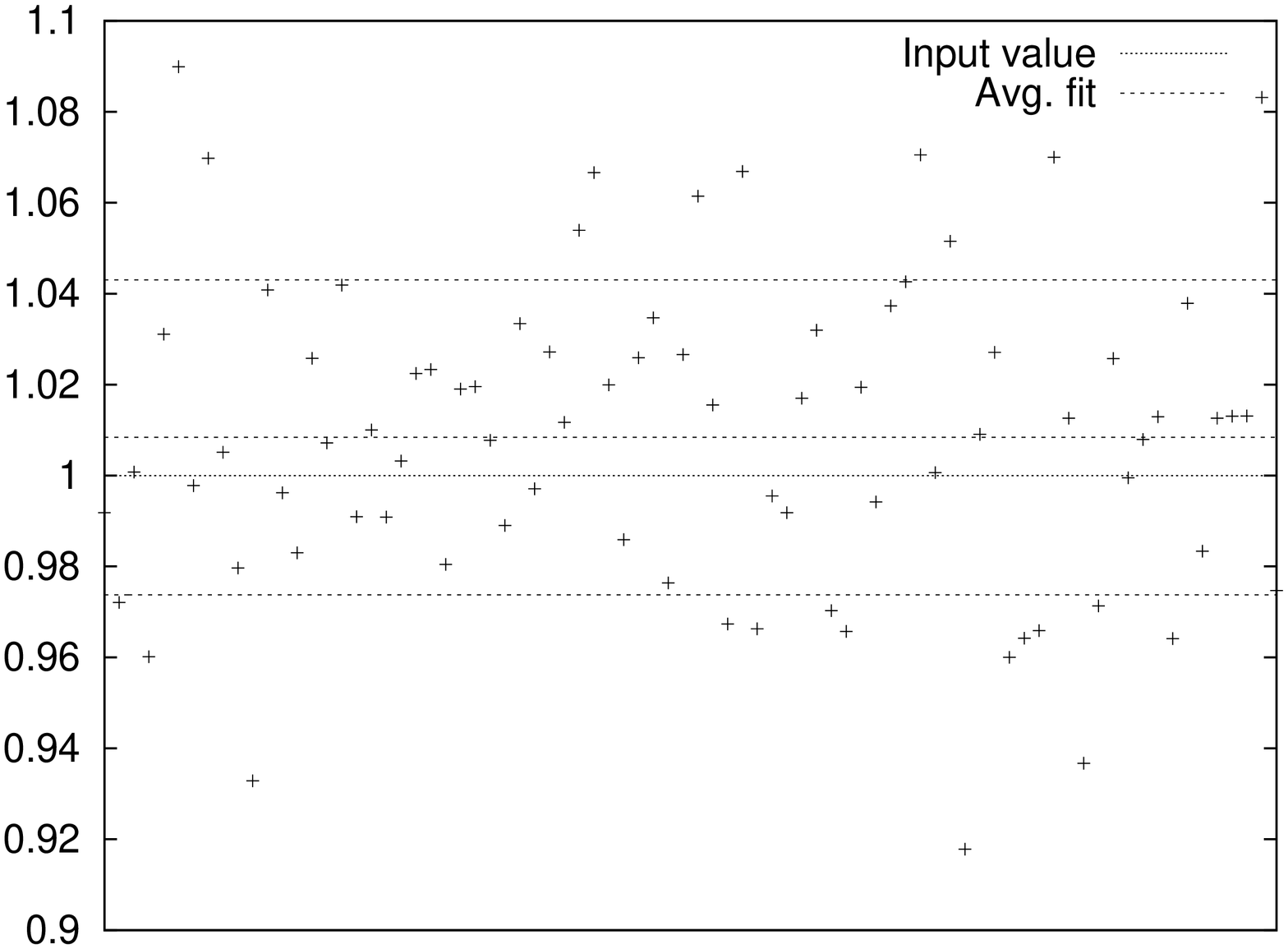} &
        \includegraphics[width=2in, angle = -90]{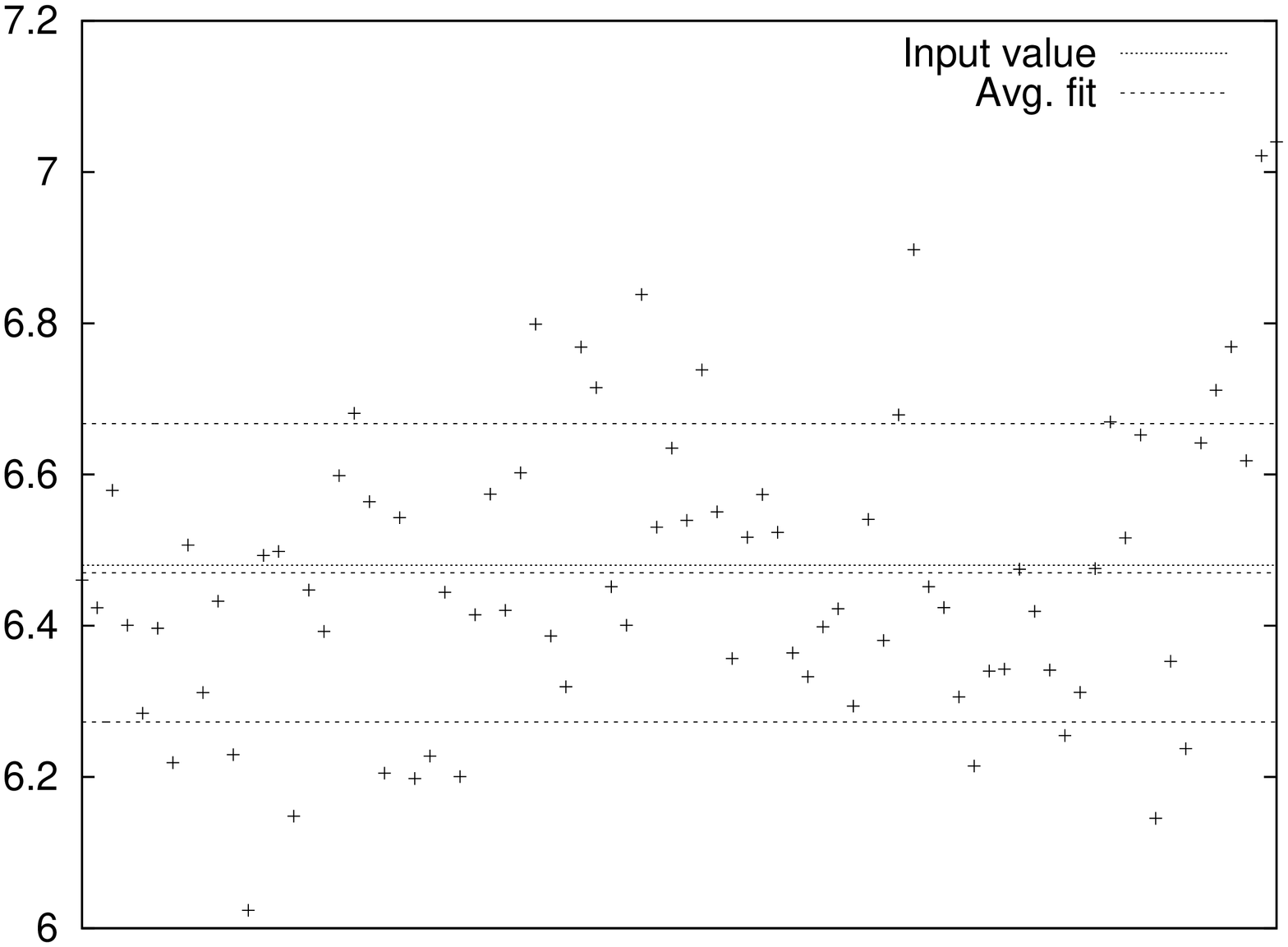}\\
        \includegraphics[width=2in, angle = -90]{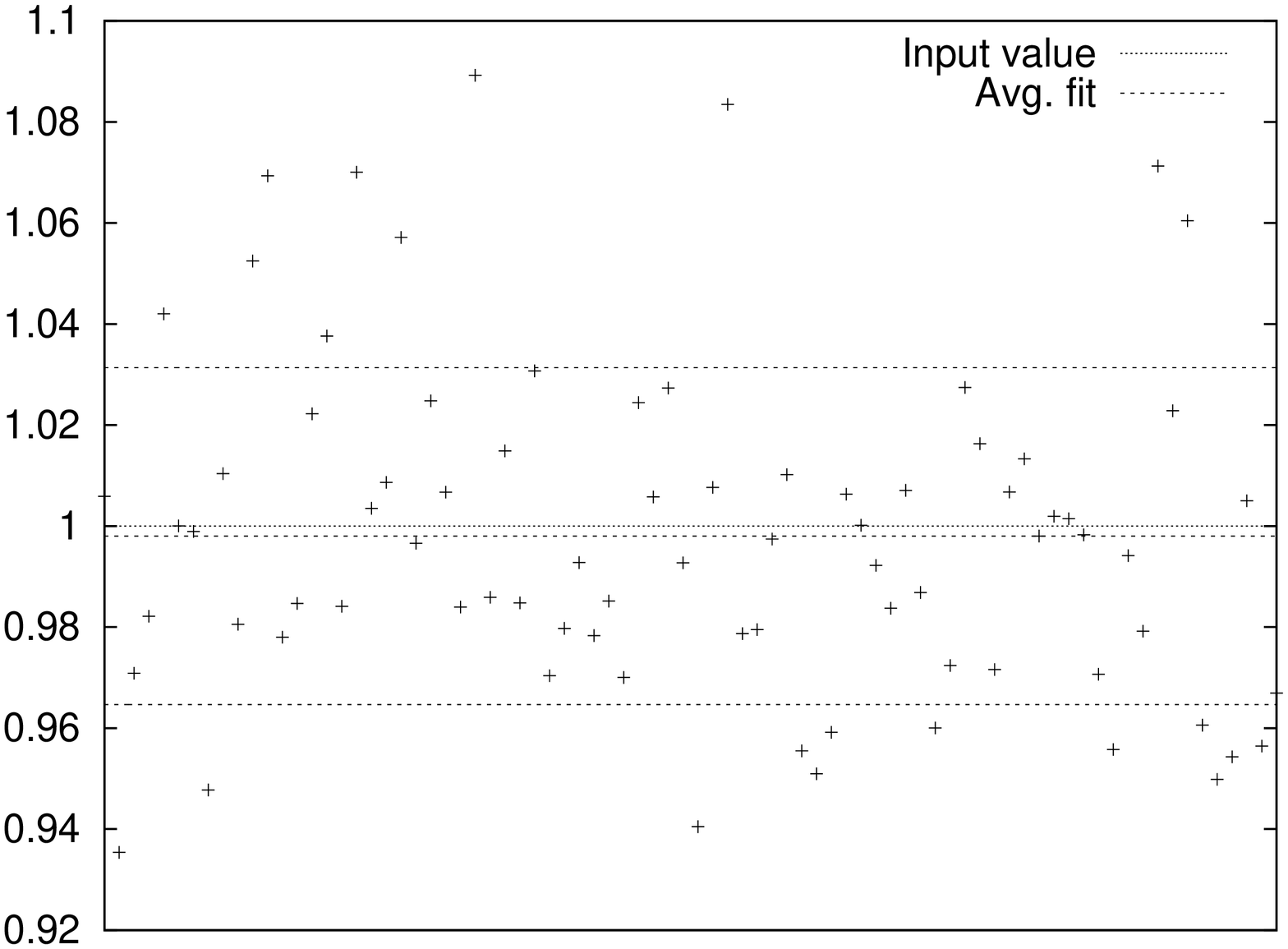} &
        \includegraphics[width=2in, angle = -90]{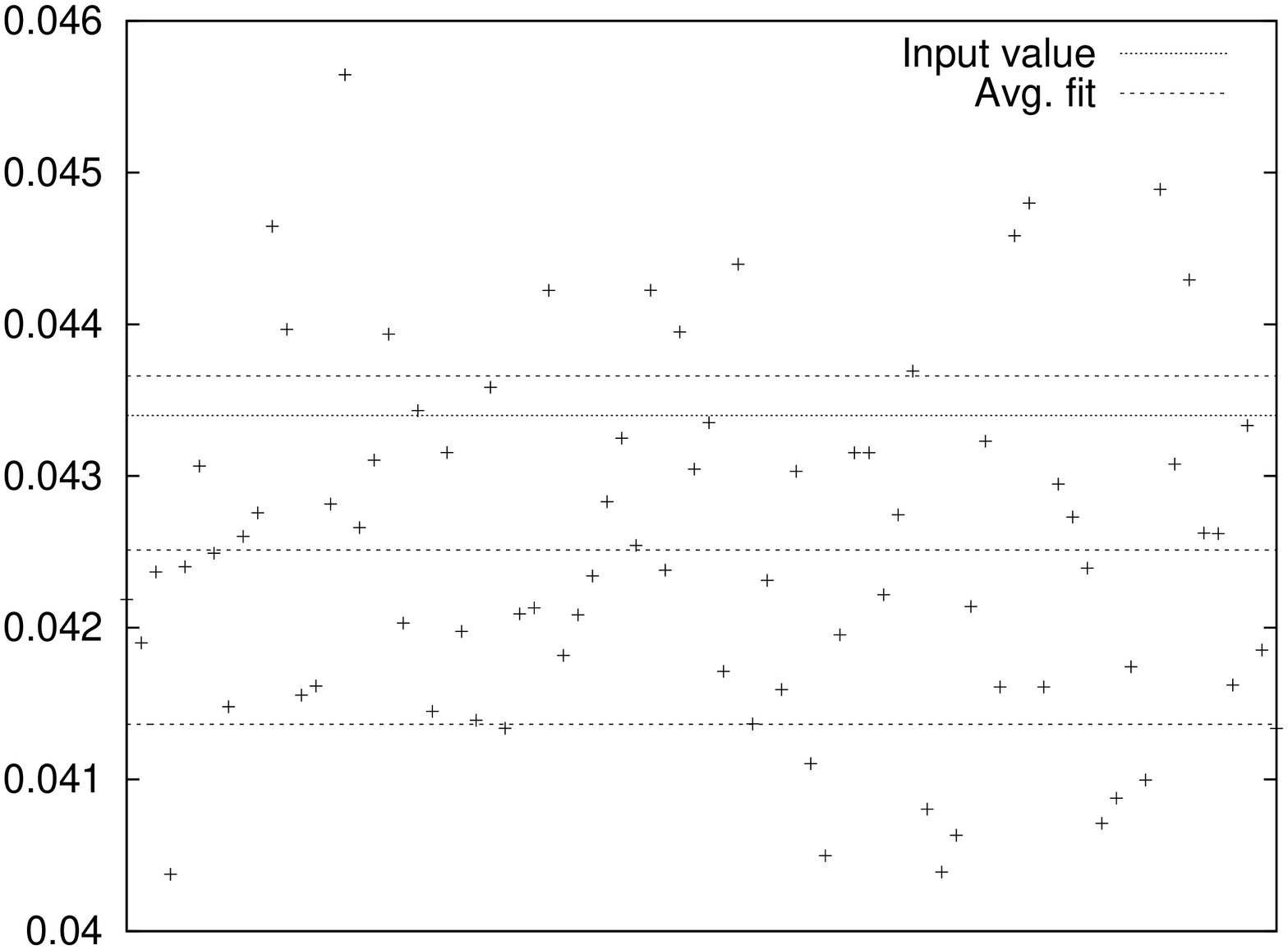}\\
    \end{tabular}
    \caption{LasDamas fitting results for the parameters $D_A^\textrm{fid}(0.35)/D_A(0.35)$ (top left), $D_A(0.35)/r_s(z_d)$ (top right), $H(0.35)/H^\textrm{fid}(0.35)$ (lower left), $H(0.35)r_s(z_d)/c$ (lower right). Dashed lines represent mean values and 1$\sigma$ error bars and input parameter values are plotted with dotted lines.}\label{ldfitplots}
\end{figure}

We use 80 LasDamas mock catalogs (1a through 40a and 1b through 40b) to validate the method discussed in section \ref{methodology}.
Each mock catalog is divided into five patches, and each patch is individually
analyzed to obtain constraints on the parameters $\{\Omega_mh^2$, $H(0.35)/H^\textrm{fid}(0.35)$, $D_A^\textrm{fid}(0.35)$$/D_A(0.35)\}$.
The estimated parameters from each mock is the weighted average of the estimates from the patches, with the weight proportional
to the galaxy count in each patch.
The parameters $\Omega_bh^2$ and $n_s$ were fixed at the simulation input values, 0.0196 and 1.0 respectively. Table \ref{ldresults} summarizes the results. All the estimated parameters are consistent within 1$\sigma$ with their input values; this provides validation of our method. We also include derived parameters $H(0.35)r_s(z_d)/c$ and $D_A(0.35)/r_s(z_d)$ as well. As shown in Fig.\ref{tiles40x40}, not all tiles are entirely full. This reduces the galaxy count in some patches and hence induces more noise compared to other patches. Therefore, we have weighted each tile appropriately before averaging and obtaining standard deviations. Fig.\ref{ldfitplots} shows the distributions of the mean values of $H(0.35)r_s(z_d)/c$ and $D_A(0.35)/r_s(z_d)$, as well as $H(0.35)r_s(z_d)/c$ and $D_A(0.35)/r_s(z_d)$,
from the 80 mocks. For reference, it also shows the standard deviation of the distributions, as well as the input values of the parameters.
\begin{table}[!h]
  \centering
    \begin{tabular}{c c c c}
        \hline
        Parameter & Mean & $\sigma$ & Input Value\\
        \hline
        $\Omega_mh^2$ & 0.1271 & 0.0049 & 0.1225\\
        $D_A^\textrm{fid}(0.35)/D_A(0.35)$ & 1.007 & 0.033 &1.0\\
        $H(0.35)/H^\textrm{fid}(0.35)$ & 1.002 & 0.035 & 1.0\\
        $D_A(0.35)/r_s(z_d)$ & 6.41 & 0.17 & 6.48\\
        $H(0.35)r_s(z_d)/c$ & 0.0425 & 0.0012 & 0.0434\\
        \hline
    \end{tabular}
    \caption{LasDamas mock catalog fitting results. Each mock catalog is divided into five patches, and each patch is analyzed separately. The estimated parameters from each mock is the weighted average of the estimates from the patches. The mean and standard deviation are obtained by averaging over 80 mock catalogs.}
\label{ldresults}
\end{table}

In order to optimize the choice for the number of patches that the survey area is divided into,
we have applied our method with different patch sizes, corresponding to 2, 5, and 10 patches respectively. Estimated parameters from the division into two patches deviate by more than 2$\sigma$ from the input values; we believe this is due to the breakdown of the flat sky approximation as each patch is about $60^\circ\times 60^\circ$. When the survey region is divided into ten patches, the number of galaxies in each patch is significantly lower and hence the power spectrum is noisy. Therefore, the covariance matrix is very noisy, and the estimated parameters have significantly larger error bars, although mean parameter values are consistent with input parameters,
as shown in Table \ref{ld10results}. We conclude that dividing the survey area into five patches is the optimal choice for this work.
\begin{table}[!h]
  \centering
    \begin{tabular}{c c c c}
        \hline
        Parameter & Mean & $\sigma$& Input Value\\\\
        \hline
        $\Omega_mh^2$ & 0.124 & 0.010  & 0.1225\\
        $D_A^\textrm{fid}(0.35)/D_A(0.35)$ & 1.017 & 0.086 &1.0\\
        $H(0.35)/H^\textrm{fid}(0.35)$ & 1.032 & 0.075 &1.0\\
        $D_A(0.35)/r_s(z_d)$ & 6.39 & 0.29 & 6.48\\
        $H(0.35)r_s(z_d)/c$ & 0.0431 & 0.0017 & 0.0434\\
        \hline
    \end{tabular}
    \caption{Same as Table \ref{ldresults}, but for dividing each mock into 10 patches.}\label{ld10results}
\end{table}

\subsection{Constraints on Parameters from SDSS Data}

\begin{figure}
    \includegraphics[width=9in, angle = 90]{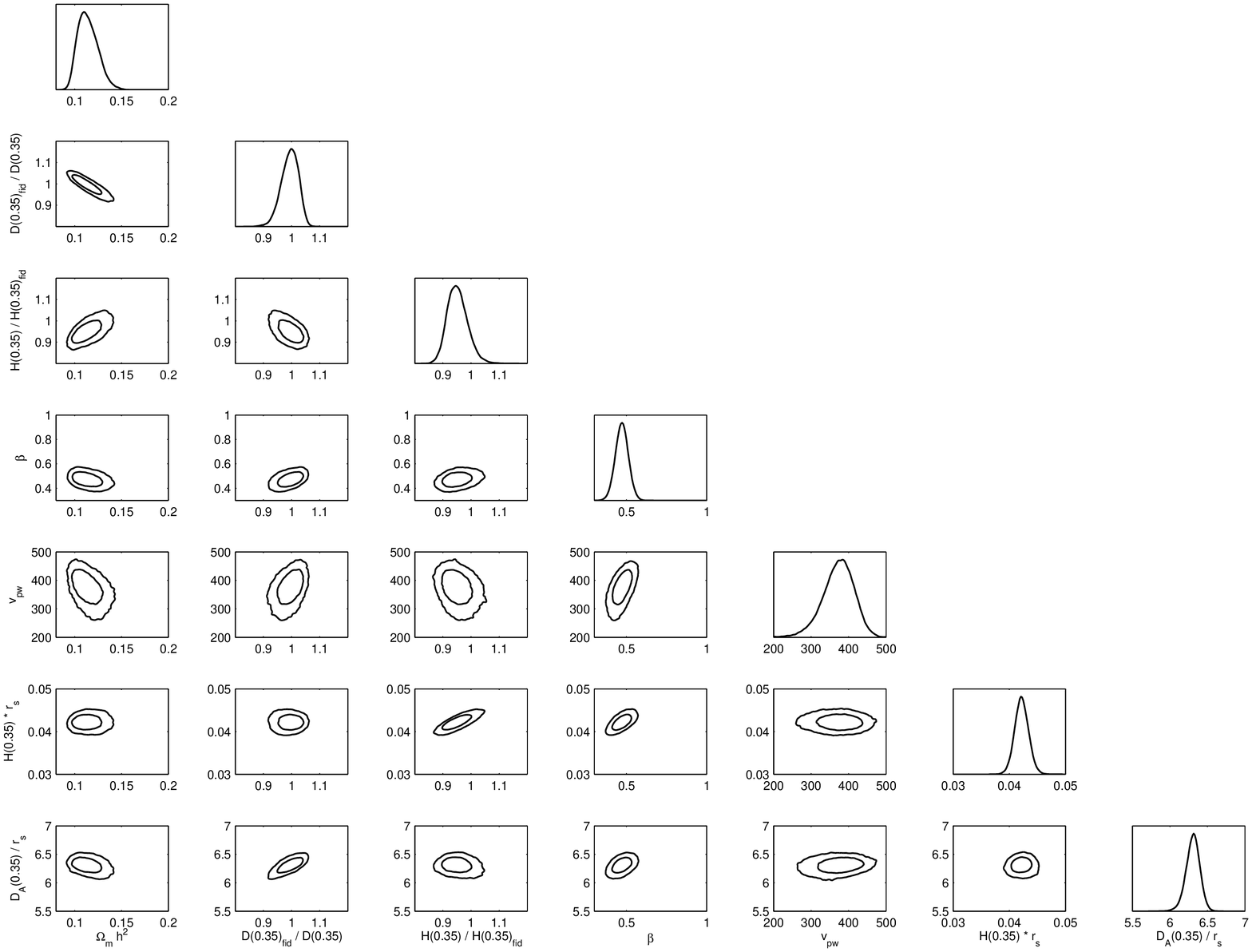}\\
    \caption{The 1D marginalized probability distribution functions and 2D joint confidence contours of the primary parameters in our analysis of SDSS DR7 LRG sample at patch 1.}\label{sdss_1of5}
\end{figure}

\begin{figure}
    \includegraphics[width=8in, angle = 90]{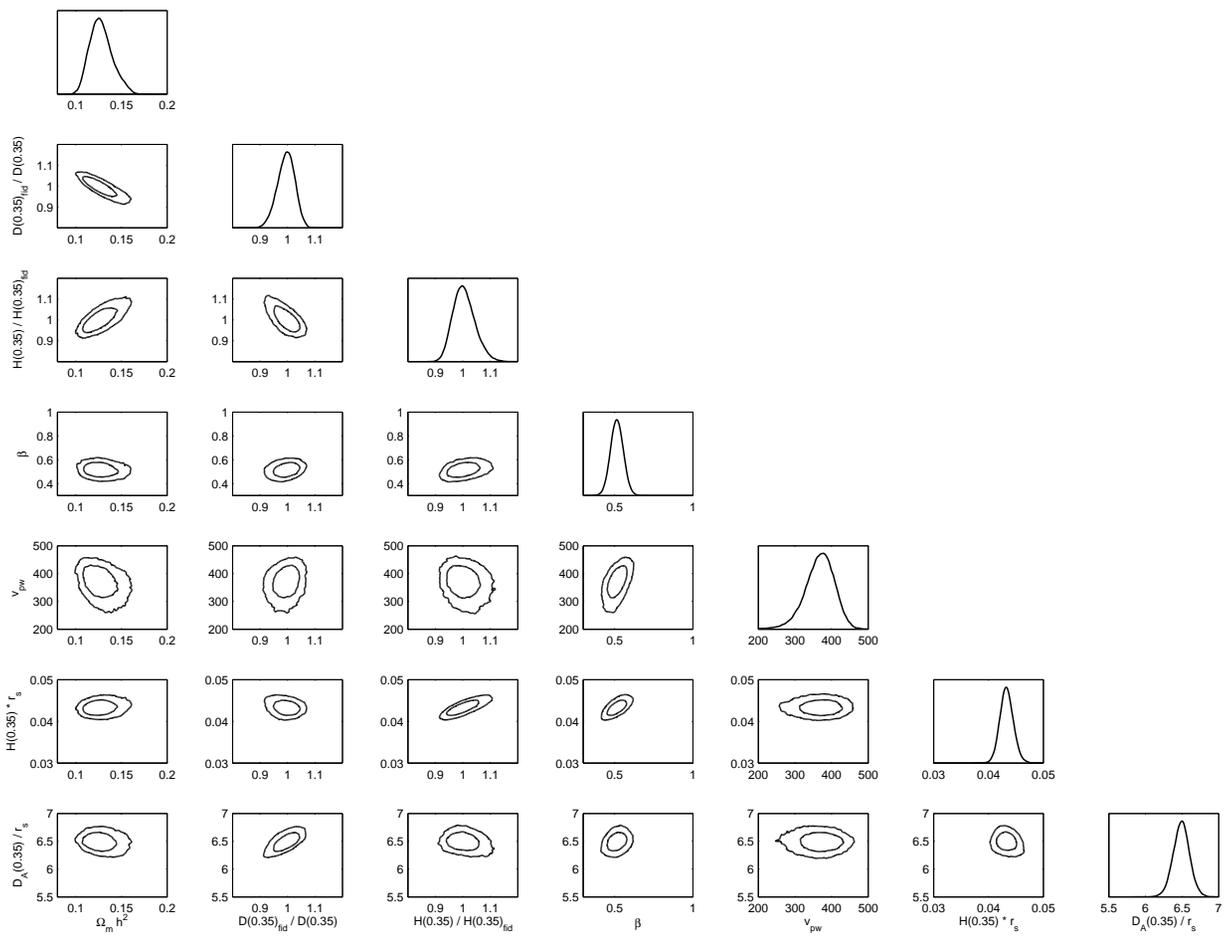}\\
    \caption{Same as Fig.\ref{sdss_1of5} for patch 2}\label{sdss_2of5}
\end{figure}

\begin{figure}
    \includegraphics[width=8in, angle = 90]{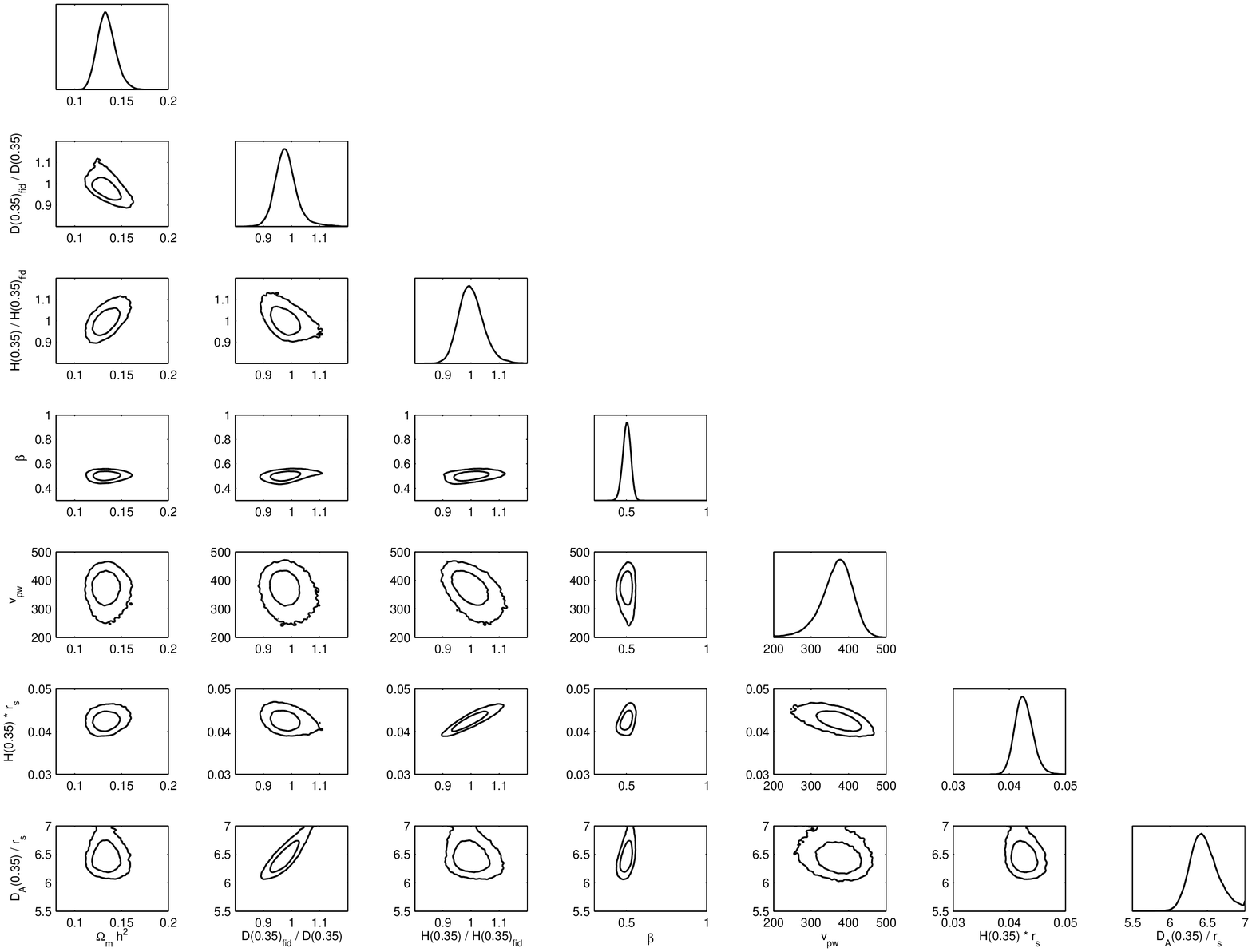}\\
    \caption{Same as Fig.\ref{sdss_1of5} for SDSS patch 3}\label{sdss_3of5}
\end{figure}

\begin{figure}
    \includegraphics[width=8in, angle = 90]{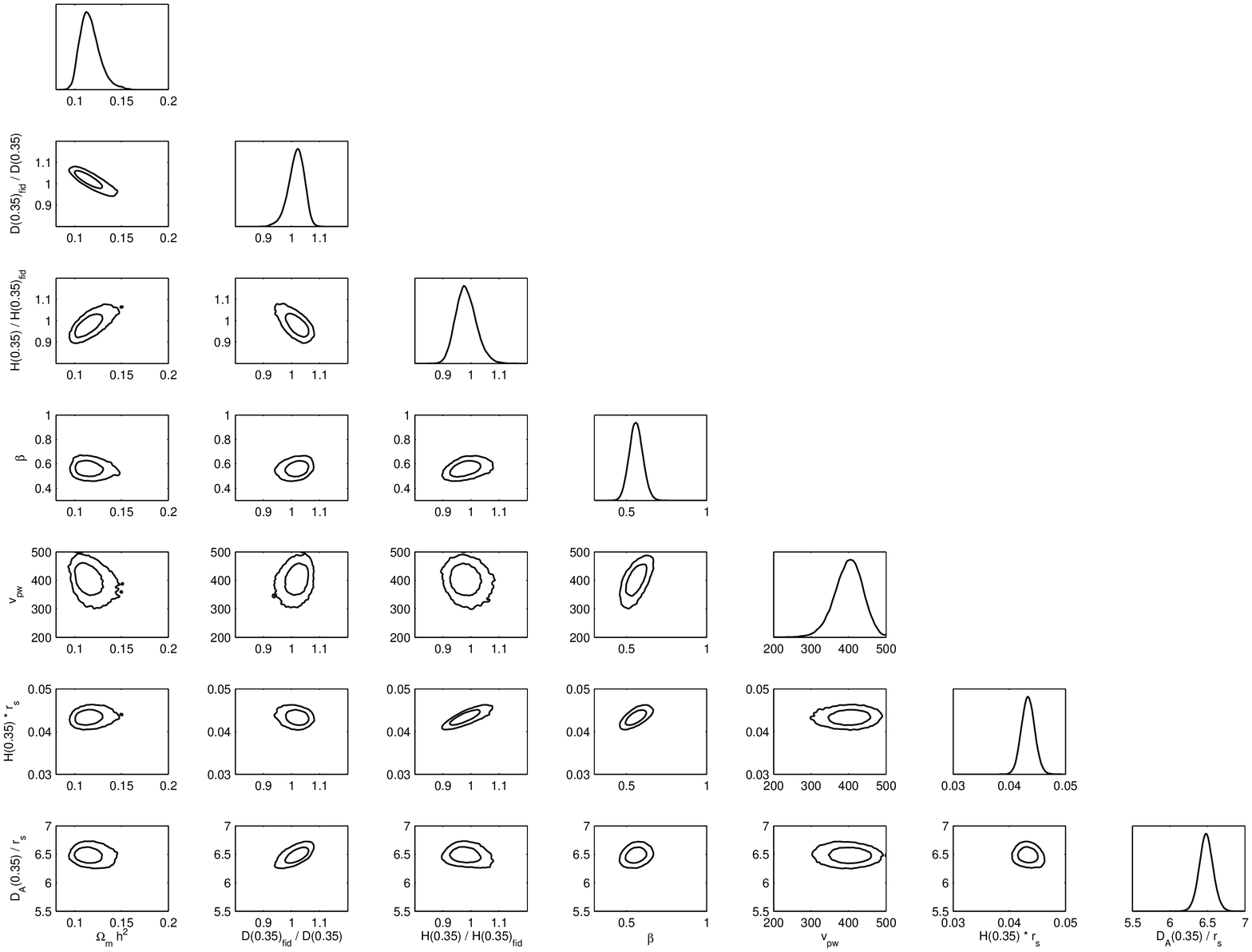}\\
    \caption{Same as Fig.\ref{sdss_1of5} for SDSS patch 4}\label{sdss_4of5}
\end{figure}

\begin{figure}
    \includegraphics[width=8in, angle = 90]{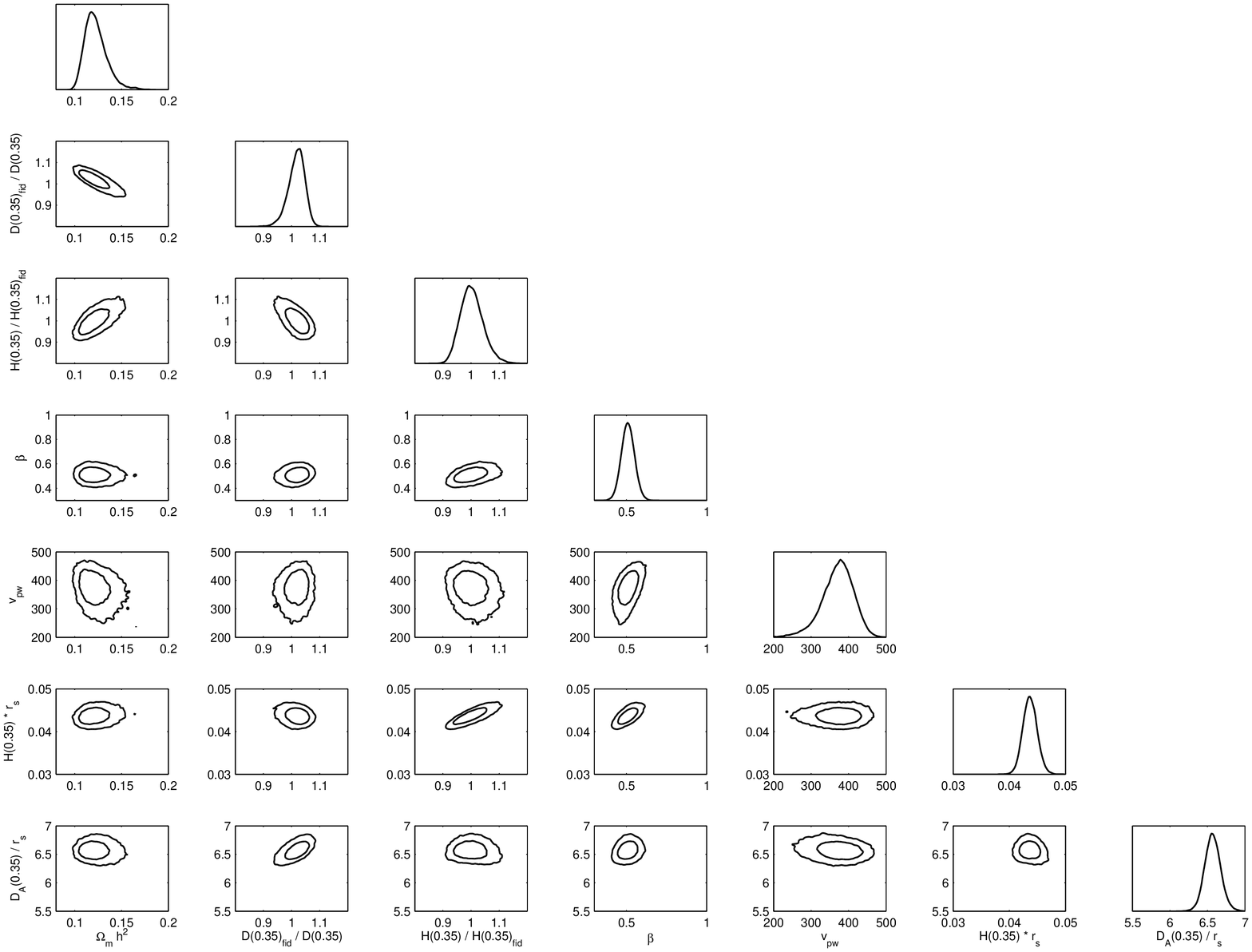}\\
    \caption{Same as Fig.\ref{sdss_1of5} for SDSS patch 5}\label{sdss_5of5}
\end{figure}

We now present our results from the analysis of SDSS DR7 LRGs. Table \ref{sdssresults} lists the mean and standard deviation
for measured parameters $\{\Omega_mh^2$, $H(0.35)$, $D_A(0.35)\}$, and derived parameters $H(0.35)r_s(z_d)/c$ and $D_A(0.35)/r_s(z_d)$ that we have obtained from the 2D power spectrum of the SDSS DR7 LRGs. The mean parameter values are calculated as follows,
\begin{equation}
p = \sum_{i=1}^5 \frac{p_i}{\sigma_i^2}\bigg{/}\sum_{i=1}^5 \frac{1}{\sigma_i^2},
\end{equation}
where, $p,p_i$ are mean parameter value and the mean parameter value for the $i^\textrm{th}$ patch, respectively.
The standard deviations are the square roots of the diagonal elements of the
covariance matrix, which is obtained by inverting the matrix sum of the inverse covariance matrices from the five patches.
Table \ref{covmats} gives the normalized covariance matrix. The covariance matrix can be reconstructed as follows:
\begin{equation}
C_{i,j}= \sigma_i \sigma_j C_{i,j}^\textrm{norm},
\end{equation}
where $C_{i,j}^\textrm{norm}$ is the normalized covariance matrix, and the $\sigma_i$'s are given in Table \ref{sdssresults}.
Figs.\ref{sdss_1of5}-\ref{sdss_5of5} show the one dimensional probability distribution functions and 2D joint confidence contours of the primary parameters in our analysis. In this analysis, we have fixed $\Omega_bh^2$ and $n_s$ to the WMAP 7 year cosmological parameter values (\cite{2011ApJS..192...16L}), 0.02258 and 0.963 respectively, and $k_{\star} = 0.11h$Mpc$^{-1}$. Fixing $\Omega_bh^2$ and $n_s$ is justified by the fact that neither parameter is well constrained by power spectrum data alone (eg. \cite{2010MNRAS.401.2148P}), and both are well determined by WMAP data. Both of these parameters were fixed in similar studies (eg. \cite{2010MNRAS.404...60R}).

\cite{2012MNRAS.426..226C} simultaneously measured $H(0.35)=82.1^{+4.8}_{-4.9}$km/s/Mpc, $D_A(0.35)=1048^{+60}_{-58}$Mpc for the first time using two-dimensional two point correlation function. Our results from using the same data set are within 1$\sigma$ of their measurements. The differences in mean values and errors can be attributed to the different methods used (correlation function versus power spectrum). Our results are also comparable with \cite{2013MNRAS.431.2834X}, where they measured $H(0.35) = 84.4\pm 7.0\,$km/s/Mpc, $D_A(0.35)=1050\pm 38\,$Mpc assuming WMAP7 cosmology from correlation function analysis of SDSS DR7 data. Their measurements are within 1$\sigma$ of our measurements. They used the multipole method to carry out an anisotropic analysis similar to \cite{2013MNRAS.431.2634C}. However, it should be noted that their theoretical model is different from Eq.(\ref{model1}): They used a different FoG model such that the denominator of Eq.(\ref{model1}) is squared. This may explain the difference in the magnitude of errors for each parameter, as the additional damping of radial power they applied is expected to result in increased uncertainty on the measured $H(z)$.
\begin{table}[!h]
  \centering
    \begin{tabular}{c c c}
        \hline
        Parameter & Mean & $\sigma$\\
        \hline
        $\Omega_mh^2$ & 0.1268 & 0.0085\\
        $D_A(0.35)$ & 1037 & 44\\
        $H(0.35)$ & 81.3 & 3.8\\
        $D_A(0.35)/r_s(z_d)$ & 6.48 & 0.25\\
        $H(0.35)r_s(z_d)/c$ & 0.0431 & 0.0018\\
        \hline
    \end{tabular}
    \caption{Results from our analysis of SDSS DR7 LRGs. The mean values and standard deviations are calculated from the  mean parameter values and covariance matrices obtained by fitting parameters for the 5 patches.}\label{sdssresults}
\end{table}

\begin{table}[!h]
    \centering
        \begin{tabular}{c c c c c}
            \hline
            $\Omega_mh^2$ & $D_A(0.35)$ & $H(0.35)$ & $D_A(0.35)/r_s(z_d)$ & $H(0.35)r_s(z_d)/c$\\
            \hline
       1   &  -0.4535   &   0.4936   &   0.1746  &  $-$0.0915\\
     $-$0.4535   &    1   &  $-$0.4009   &  -0.2772  &    0.9270\\
      0.4936   &  $-$0.4009    &   1   &   0.9420  &   $-$0.2435\\
      0.1746   &  $-$0.2772    &  0.9420    &   1  &   $-$0.2384\\
    $-$0.0915   &   0.9270    & $-$0.2435    & $-$0.2384   &    1\\
            \hline
    \end{tabular}
    \caption{Normalized average covariance matrix corresponding to Table \ref{sdssresults}.}\label{covmats}
\end{table}

\section{Conclusion and Discussion}\label{discussion}
We present the first measurement of $H(z)$ and $D_A(z)$ from the two-dimensional galaxy power spectrum from SDSS DR7 LRG data. This method can be applied to any future survey with a broad sky coverage. The basic concept is to divide the sky into patches of roughly equal area and calculate individual power spectra for each patch. We find that the optimum number of patches for SDSS DR7 data is five, so that enough number of galaxies are included in each patch and the flat sky approximation is also valid. We have measured $\{\Omega_mh^2$, $H(0.35)$, $D_A(0.35)\}$ and derived parameters $H(0.35)r_s(z_d)/c$ and $D_A(0.35)/r_s(z_d)$ from the SDSS DR7 LRGs, as shown in Table \ref{sdssresults}. Note that we have analyzed the full two-dimensional power spectrum, and not the Baryon Acoustic Oscillation (BAO) alone.

To validate our method, we applied it to LasDamas mock data and constrained cosmological parameters. The results shown in Table \ref{ldresults} are consistent with the LasDamas input parameters, thus establishing the validity of our method.

Our measurements of $H(0.35)$ and $D_A(0.35)$ from the SDSS DR7 LRGs, with errors
of 4.67\% and 4.29\% respectively, are comparable with the values reported in similar work.
We also find that the derived parameters $H(0.35)r_s(z_d)/c$ and $D_A(0.35)/r_s(z_d)$ are more tightly constrained, with errors of 4.18\% and 3.87\% respectively.
A survey such as BOSS which is currently ongoing with more galaxies and deeper than SDSS would enable the utilization of this method to further tighten the constraints on these parameters, as well as the matter density and index of the primordial power spectrum.

\section{ACKNOWLEDGEMENTS}
MDPH thanks Will Percival, Gert H{\"u}tsi, and Chris Blake for valuable discussions.  We are grateful to the LasDamas project for making their mock catalogues publicly available. Computational facilities for this project were provided by the OU Supercomputing Center for Education and Research (OSCER) at the University of Oklahoma (OU) and we thank OSCER Director Henry Neeman for invaluable technical support. This work was supported in part by DOE grants DE-FG02-04ER41305 and DE-SC0009956. C.C. was also supported by the Spanish MICINN's Consolider-Ingenio 2010 Programme under grant MultiDark CSD2009-00064 and grant AYA2010-21231, and by the Comunidad de Madrid under grant HEPHACOS S2009/ESP-1473.

Funding for the Sloan Digital Sky Survey (SDSS) has been provided by the Alfred P. Sloan Foundation, the Participating Institutions, the National Aeronautics and Space Administration, the National Science Fondation, the U. S. Department of Energy, the Japanese Monbukagakusho, and the Max Planck Society. The SDSS Web site is \url{http://www.sdss.org/}.

The SDSS is managed by the Astrophysical Research Consortium (ARC) for the Participating Institutions. The Participating Institutions are The University of Chicago, Fermilab, the Institute for Advanced Study, the Japan Participation Group, The Johns Hopkins University, Los Alamos National Laboratory, the Max-Planck-Institute for Astronomy (MPIA), the Max-Planck-Institute for Astrophysics (MPA), New Mexico State University, University of Pittsburgh, Princeton University, the United States Naval Observatory, and the University of Washington.


\newpage
\bibliographystyle{mn2e}
\bibliography{HzDz2DPk}

\end{document}